\documentclass[twocolumn,english,pra,showpacs,aps]{revtex4-1}
\usepackage[T1]{fontenc}
\usepackage[latin9]{inputenc}
\usepackage[nointlimits]{amsmath}
\usepackage{amssymb}
\usepackage{graphicx}
\usepackage{esint}

\usepackage{babel}
 \usepackage{color}

\makeatletter
 
 \@ifundefined{textcolor}{}
 {%
   \definecolor{BLACK}{gray}{0}
   \definecolor{WHITE}{gray}{1}
   \definecolor{RED}{rgb}{1,0,0}
   \definecolor{GREEN}{rgb}{0,1,0}
   \definecolor{BLUE}{rgb}{0,0,1}
   \definecolor{CYAN}{cmyk}{1,0,0,0}
   \definecolor{MAGENTA}{cmyk}{0,1,0,0}
   \definecolor{YELLOW}{cmyk}{0,0,1,0}
 }

\begin{document}

\title{Sensitivity to thermal noise of atomic Einstein-Podolsky-Rosen
entanglement}

\author{R.~J.~Lewis-Swan}

\author{K.~V.~Kheruntsyan}

\affiliation{The University of Queensland, School of Mathematics and Physics,
Brisbane,
Qld 4072, Australia}

\begin{abstract}
We examine the prospect of demonstrating Einstein-Podolsky-Rosen (EPR) 
entanglement for massive particles using spin-changing collisions in a spinor
Bose-Einstein condensate. 
Such a demonstration has recently been attempted by Gross
\textit{et al.} [Nature \textbf{480}, 219 (2011)] using a condensate of
$^{\text{87}}$Rb atoms trapped 
in an optical lattice potential. For the condensate initially prepared in 
the $(F,m_{F})=(2,0)$ hyperfine state, with no population in the
$m_{F}=\pm1$ states, we predict a significant suppression of the
product of inferred quadrature variances below the Heisenberg uncertainty
limit, implying strong EPR entanglement. However, such EPR entanglement is lost when the collisions are initiated
in the presence of a small (currently undetectable) thermal population
$\bar{n}_{\mathrm{th}}$
in the $m_{F}=\pm1$ states. For condensates containing $150$ to $200$ atoms, we
predict an upper bound
of $\bar{n}_{\mathrm{th}}\simeq 1$ that can be tolerated in this experiment
before EPR entanglement is lost. \end{abstract}


\date{\today}

\maketitle

\section{Introduction}

Entanglement has proven to be ``the characteristic trait of quantum
mechanics'' as first coined by Schr\"odinger \cite{CambridgeJournals:1737068}.
It
forms the foundations
of quantum information theory and quantum computing. Further, in interferometry
entanglement enables measurement precision to surpass the standard quantum
limit \cite{Giovannetti19112004}. This is particularly important in
atom interferometry \cite{Oberthaler-interferometer,*Riedel-interferometer} as
atom flux is generally limited.
However, the most important foundational trait of entanglement comes with 
its role in the Einstein-Podolsky-Rosen paradox (EPR) \cite{EPR:35,*Bohr:35}.
This requires the underlying quantum correlations to 
be stronger than those satisfying the simpler inseparability criteria. The resulting EPR-entanglement criterion
confronts the Heisenberg uncertainty relation and 
puts us into the context of EPR arguments that question the
completeness of quantum mechanics and open the door to alternative descriptions of these correlations via local hidden variable theories \cite{Bohm:52,*Bohm-Aharonov,*Bell}.
The
EPR paradox for continuous-variable quadrature observables
\cite{PhysRevA.40.913}
(which are analogous to the position and momentum observables originally
discussed by EPR) has been demonstrated in optical 
parametric down-conversion \cite{PhysRevLett.68.3663} and most recently attempts
have been made
to demonstrate \cite{Oberthaler-2011} the paradox with ensembles of massive particles generated 
by spin-changing collisions in a spinor Bose-Einstein condensate (BEC)
\cite{PhysRevLett.85.3987,PhysRevLett.85.3991}.

In this paper, we seek to provide a theoretical treatment of the recent
experiment
by Gross \textit{et al.} \cite{Oberthaler-2011} which reported entanglement, or
quantum inseparability, 
of two atomic ensembles produced by spin-changing collisions in a $^{87}$Rb BEC.
For the BEC initially prepared in the 
$(F,m_F)=(2,0)$ hyperfine state, the collisions produce correlated 
pairs of atoms in the $m_F=\pm 1$ sublevels. The authors observed that the
resulting state 
was inseparable, though a measurement of a stronger EPR entanglement criterion
was inconclusive.
A normalized product of inferred quadrature variances of $4\pm17$
was reported, whereas a demonstration of the EPR paradox requires this quantity
to be less than unity
\cite{PhysRevA.40.913, PhysRevA.78.060104}.

The short-time dynamics of the spin-mixing process, for a vacuum initial state
of the $m_{F}=\pm1$ atoms, is similar to that of a spontaneous 
parametric down-conversion in the undepleted pump approximation. This 
paradigmatic nonlinear optical process is known to produce an EPR entangled
twin-photon state that 
can seemingly violate the Heisenberg uncertainty relation 
for inferred optical quadratures~\cite{PhysRevA.40.913}. 
Such a violation has previously been observed in 1992 by Ou \textit{et
al.}~\cite{PhysRevLett.68.3663}.
Due to the inconclusive nature of an analogous measurement of matter-wave
quadratures 
in Ref. \cite{Oberthaler-2011}, we seek to perform a theoretical
analysis of spin-changing dynamics 
and calculate various measures of entanglement in experimentally realistic
regimes. 
In particular, we focus on the sensitivity of 
EPR entanglement to an initial population in the $m_{F}=\pm1$ sublevels with
thermal statistics. 
In the optical case this question is argued to be irrelevant as
at optical frequencies and room temperatures the thermal population of the
signal and idler modes 
is negligible, allowing us to safely approximate them as vacuum states. However, 
these considerations are inapplicable to ultracold atomic gases. This was highlighted recently by Mel\'e-Messeguer \textit{et. al.} \cite{Santos-thermal}, who 
quantitatively predicted the possibility of non-trivial thermal activation of the $m_{F}=\pm1$ sublevels in a spin-1 BEC.
 Accordingly, when interpreting experimental results care must be taken 
in differentiating spin-mixing dynamics initiated by vacuum noise from that initiated by  
thermal noise or a small coherent seed
\cite{PhysRevLett.104.195303}.
To this end, our modelling of the experiment of Gross \textit{et al.}
\cite{Oberthaler-2011} 
is more consistent with a small thermal population in the $m_{F}=\pm1$ 
sublevels, rather than a vacuum initial state or small coherent seed.
From a broader perspective, the connection between our
results and 
the widely applicable model of parametric down-conversion highlights the generally fragile nature of 
atomic EPR entanglement to thermal noise, demonstrating that future experiments
must be refined to
overcome this problem.

\section{The system}

The experiment of Ref. \cite{Oberthaler-2011} starts with
a BEC of $^{\text{87}}$Rb atoms prepared in the $(F,m_{F})=(2,0)$ state and
trapped
in a one-dimensional optical lattice. The lattice potential is sufficiently
deep to prevent tunnelling between neighbouring wells. Furthermore, due to the relatively small number of atoms in each well, the spin-healing length 
is of the order of the spatial size of the condensate 
in the well meaning the spatial dynamics of the system are frozen, and hence we may treat
the condensate in each well according to the single-mode approximation \cite{PhysRevLett.81.5257, PhysRevA.60.1463, chang2005coherent}. In this approximation the field operator
$\hat{\psi}_{i}(\mathbf{r})$ for each component $i\!\equiv \!m_F\!=\!0,\pm1,\pm2$
is expanded as $\hat{\psi}_{i}(\mathbf{r})\!=\!\phi(\mathbf{r})\hat{a}_{i}$,
where $\phi(\mathbf{r})$ is the common spatial ground state wavefunction ($\phi_{i}(\mathbf{r})\equiv\phi(\mathbf{r})$) and $\hat{a}_{i}$ is the respective bosonic annihilation operator.

A quadratic Zeeman shift and microwave dressing of the $m_{F}\!=\!0$
state is employed to energetically restrict the spin-mixing dynamics
to the $m_{F}\!=\!0,\pm1$ states \cite{Oberthaler-2011}, and
so 
for short time durations we may map the spin-2 system to an effective spin-1
Hamiltonian
\cite{Kawaguchi2012253}
of the form
$\hat{H}=\hat{H}_{\mathrm{inel}}+\hat{H}_{\mathrm{el}}+\hat{H}_{\mathrm{Z}}$, 
\begin{eqnarray}
\hat{H}_{\mathrm{inel}}  & = &
\hbar
g(\hat{a}_{0}^{\dagger}\hat{a}_{0}^{\dagger}\hat{a}_{-1}\hat{a}_{1}+\hat{a}_{1}^{\dagger}\hat{a}_{-1}^{\dagger}
\hat{a}_{0}\hat{a}_{0}),\\
\hat{H}_{\mathrm{el}}  & = & \hbar g(\hat{n}_0\hat{n}_{1}+\hat{n}_0\hat{n}_{-1}) \\
\hat{H}_{\mathrm{Z}}  & = & -p\left(\hat{n}_{1}-\hat{n}_{-1}\right)-q\left(\hat{n}_{1}+\hat{n}_{-1}\right)
\end{eqnarray}
where $\hat{n}_{i}=\hat{a}^{\dagger}_{i}\hat{a}_{i}$ is the particle number
operator and $i\!=\!0,\pm1$ are referred to, respectively, as the pump and
signal/idler modes from
herein. We have ignored terms proportional to $\hat{N}(\hat{N}-1)$ in $\hat{H}$ as this is a conserved quantity and contributes only a global phase rotation. 
The inelastic spin-changing collisions are described by
$\hat{H}_{\mathrm{inel}}$, and  
the remaining elastic $s$-wave scattering terms are grouped in
$\hat{H}_{\mathrm{el}}$,
where $g$ is the coupling constant associated with $s$-wave collisions
\cite{Kawaguchi2012253}. For a spin-2 system, the coupling 
is given by $g=\frac{6}{14}(3g_{4}+4g_{2})\int
d^{3}\mathbf{r}\left|\phi(\mathbf{r})\right|^{4}$, where
 $g_{F}=4\pi\hbar^{2}a_{F}/m$ describes
$s$-wave scattering with total spin $F$, characterised by scattering
length $a_{F}$ \cite{Kawaguchi2012253}. For comparison, for an actual spin-1 system the coupling constant 
would be given by $g=\frac{g_{2}-g_{0}}{3}\int
d^{3}\mathbf{r}\left|\phi(\mathbf{r})\right|^{4}$, where
 $g_{F}=4\pi\hbar^{2}a_{F}/m$.
In our representation of
$\hat{H}_{\mathrm{el}}$ we have used the fact that the relative number difference, $\hat{n}_{1} - \hat{n}_{-1}$, is a conserved quantity. The interaction with the
magnetic field is described by $\hat{H}_{\mathrm{Z}}$, 
where the
linear and quadratic Zeeman effects are parametrized, respectively, by
$p=\mu_{B}B_{0}/2$ and $q=p^{2}/\hbar\omega_{\mathrm{HFS}}$
\cite{PhysRevA.72.063619}, with 
$\omega_{\mathrm{HFS}}/2\pi\approx6.835$ GHz being
the hyperfine splitting frequency of $^{87}$Rb \cite{Salomon-Rb87-EPL-1999}.
For our initial conditions the relative number difference, $\hat{n}_{1} - \hat{n}_{-1}$, will always be zero and hence we may ignore the linear Zeeman effect. 
We may also redefine the parameter $q$ to absorb the effects of microwave level dressing (used by Gross \textit{et al.} \cite{Oberthaler-2011}) and any other fixed
energy shift between the $m_F=0$ and $m_F=\pm1$ energy levels.

Simple analogies between the states of the
signal and idler modes in spin-changing
collisions and optical parametric down-conversion consider
only $\hat{H}_{\mathrm{inel}}$ in the undepleted pump approximation, however,
competing mean-field
($\hat{H}_{\mathrm{el}}$) and Zeeman
($\hat{H}_{\mathrm{Z}}$) effects lead to additional dynamics
\cite{PhysRevLett.97.110404} due to dephasing. 
The full Heisenberg operator equations of motion are
given by 
\begin{eqnarray}
\frac{d\hat{a}_{0}}{d\tau} & = &
-i\left[2\hat{a}_{-1}\hat{a}_{1}\hat{a}_{0}^{\dagger}+\left(\hat{n}_{1}
+\hat{n}_{-1}\right)\hat{a}_{0}\right],
\label{eq:da0_dt}\\
\frac{d\hat{a}_{\pm1}}{d\tau} & = &
-i\left[\hat{a}_{0}^{2}\hat{a}_{\mp1}^{\dagger}+\left(\hat{n}_{0} -q/g
\right)\hat{a}_{\pm1}\right],\label{eq:da1_dt}
\end{eqnarray} 
where we have introduced $\tau=gt$ as dimensionless time. We see that the phase
accrued in the $\hat{a}_{\pm1}$ modes grows
$\propto\left(\hat{n}_{0} -q/g\right)$
whilst for the $\hat{a}_{0}$ mode the phase grows
$\propto\left(\hat{n}_{1}+\hat{n}_{-1}\right)$.
In the short-time undepleted pump approximation \cite{walls2010quantum},
this is equivalent to a phase rotation
$\hat{a}_{\pm1}\rightarrow\hat{a}_{\pm1}\mathrm{exp}[i\left(N_{0} -q/g\right)\tau
]$, where $N_{0}=\langle\hat{n}_{0}(0)\rangle$ is the initial population
of the $m_{F}=0$ component. This rotation leads to a dynamical 
phase mismatch between
the spinor components that decelerates the pair-production process
\cite{PhysRevLett.97.110404}.
To prevent phase mismatch in the short-time limit one can 
choose $q\!=\!gN_{0}$ in which case Eqs. (\ref{eq:da0_dt})-(\ref{eq:da1_dt}) reduce to
those of resonant down-conversion \cite{walls2010quantum}.

\section{Results and discussion}

\subsection{Population dynamics}

\begin{figure}
\includegraphics[width=8.2cm]{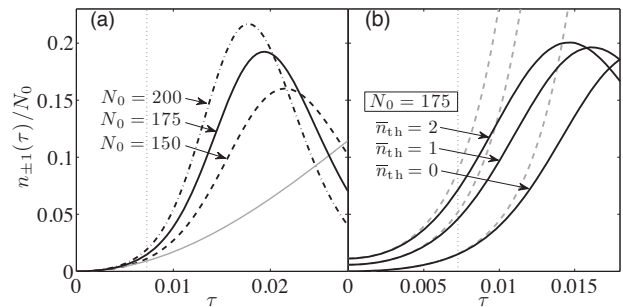}
\caption{(a) Fractional population $n_{\pm1}(\tau)/N_0$ of the 
signal/idler modes [where $n_{\pm1}(\tau)\!\equiv \!\langle
\hat{a}^{\dagger}_{\pm1}(\tau)\hat{a}_{\pm1}(\tau)
\rangle$]
as a function of the dimensionless time $\tau$, for vacuum initial state and
different initial number of atoms in the pump mode, $N_0$. 
The quadratic Zeeman term is phase-matched to $q\!=g\!N_0$ in all cases, except for
the grey solid line which is shown for comparison for $q\!=\!0$ and $N_0\!=\!175$. 
The vertical dotted line indicates the 
measurement time $\tau'\!=\!0.0073$ used in Ref. \cite{Oberthaler-2011}. (b)
Same as in (a) but with thermally seeded populations in the signal/idler
modes (assumed to be equal to each other), for $N_0\!=\!175$.
The grey dashed lines show the analytic predictions in the 
undepleted pump approximation. 
\label{fig:spin-mixing}}
\end{figure}

We first analyze the spin-changing dynamics for the case of a vacuum initial 
state for the signal/idler modes, and a coherent state
$|\alpha_{0}(0)\rangle$ for the pump mode 
with initial number of atoms $N_0=|\alpha_0(0)|^2$. 
This case can be treated in a straightforward manner (see, e.g., Ref.
\cite{PhysRevA.78.060104}) 
by diagonalizing the full Hamiltonian in the truncated Fock-state basis and
solving the 
Schr\"odinger equation (see also \cite{coherent}). 
Figure \ref{fig:spin-mixing}
(a) shows the population dynamics of the signal and idler 
modes, for different initial atom numbers $N_0$ and the quadratic Zeeman 
term tuned to the phase-matching condition $q=gN_0$. Setting $q=0$ eliminates
the Zeeman shift and we observe (grey solid line) significantly slowed dynamics due
to phase mismatch. For reference,
we also mark the experimental measurement time of Ref. \cite{Oberthaler-2011},
$\tau'=0.0073$, corresponding to the reported value of the squeezing parameter
$r\equiv N_0\tau'\simeq2$ 
\cite{walls2010quantum}, evaluated for $N_0=275$.

We next analyze the case of an initial thermal seed
in the signal/idler modes, with an equal average number of 
atoms $\bar{n}_{\mathrm{th}}$ in both modes. To simulate the dynamics in this
case, we 
use the truncated Wigner method (Ref.~\cite{Olsen20093924} gives 
simple prescriptions on how to model various initial states in the Wigner
representation). Figure~\ref{fig:spin-mixing}~(b) illustrates that
the presence of the thermal seed accelerates population growth, however, 
it does not significantly effect the maximal depletion of the
BEC. The numerical results in Fig.~\ref{fig:spin-mixing}~(b) are 
compared with the analytic predictions of the simple model of parametric 
down-conversion in the undepleted pump approximation, 
$n_{\pm1}(\tau)\!=\!\sinh^2(N_0\tau)[1+2\bar{n}_{\mathrm{th}}]+\bar{n}_{\mathrm{th}}$ (see Appendix A for full analytic solutions). 
As expected, we find good agreement between the numerical and analytic results
in the short-time limit. We also conclude that as far as the mode populations are concerned,
the experimental measurement time
$\tau'\!=\!0.0073$ 
is not too far from the regime of validity of the simple analytic model, 
at least for $\bar{n}_{\mathrm{th}}\! \lesssim \!2$. This conclusion, 
however, cannot neccesarily be carried through to other observables, 
such as entanglement measures analysed below.

\subsection{EPR entanglement}

Central to this paper is an investigation into the possible demonstration
of the EPR paradox as outlined in Ref.
\cite{Oberthaler-2011}.
In the context of continuous-variable entanglement,
this is equivalent to the seeming violation of the Heisenberg uncertainty
relation
for inferred quadrature variances \cite{PhysRevA.40.913,PhysRevA.78.060104}. In
the normalised form this EPR entanglement criterion can be written as
\begin{equation}
\Upsilon_j=\frac{\Delta_{\mathrm{inf}}^{2}\hat{X}_{j}\Delta_{\mathrm{inf}}^{2}
\hat{Y}_{j}}{(1-\langle \hat{a}_{j}^{\dagger}\hat{a}_{j}\rangle
/\langle \hat{b}_{j}^{\dagger}\hat{b}_{j}\rangle)^2}<1,
\label{EPR-criterion}
\end{equation}
where the optimal \footnote{The form of the inferred quadrature variance in Eq.~(\ref{inf_quad_var}) varies slightly from that used in Ref. \cite{Oberthaler-2011}, where the inferred quadrature 
variances are equivalent to measurements of $\Delta^2_{\mathrm{inf}}\hat{X}_2=\Delta^{2}(\hat{X}_1 - \hat{X}_2)$ and $\Delta^2_{\mathrm{inf}}\hat{Y}_2=\Delta^{2}(\hat{Y}_1 + \hat{Y}_2)$ \cite{PhysRevA.78.060104}. 
This choice is different in that it does not give the optimal violation 
of Eq.~(\ref{EPR-criterion}) \cite{PhysRevA.40.913}, however, in the parameter regime we consider the difference between the choices of inferred quadratures is 
not qualitatively significant.} inferred quadrature variance for $\hat{X}_j$ (and similarly for
$\hat{Y}_j$) is given by \cite{PhysRevA.40.913}
\begin{equation}
\Delta_{\mathrm{inf}}^{2}\hat{X}_{j} = \langle
(\Delta\hat{X}_{j})^{2}\rangle -\frac{\langle
\Delta\hat{X}_{i}\Delta\hat{X}_{j}\rangle ^{2}}{\langle
(\Delta\hat{X}_{i})^{2}\rangle },
\label{inf_quad_var}
\end{equation}
with $\Delta\hat{X}_{j}\equiv\hat{X}_{j}-\langle
\hat{X}_{j}\rangle$ and $i,j=\pm1$. The generalized quadrature operators 
are defined as 
$\hat{X}_{j}\left(\theta\right)=(\hat{a}_{j}^{\dagger}\hat{b}_{j}e^{i\theta}
+\hat{b}_{j}^{\dagger}\hat{a}_{j}e^{-i\theta})/\langle\hat{b}_{j}^{\dagger}\hat{
b}_{j}\rangle^{1/2}$ \cite{PhysRevA.78.060104}, where the operator $\hat{b}_{j}$ represents the local oscillator field required for homodyne detection 
of the quadratures and we denote $\hat{X}_{j}\!=\!\hat{X}_{j}\left(\pi/4\right)$ and
$\hat{Y}_{j}\!=\!\hat{X}_{j}\left(3\pi/2\right)$. Choosing this pair of canonically conjugate quadratures maximises the correlation (anti-correlation)
between them, defined as 
$C = \langle \hat{X}_i(\theta)\hat{X}_j(\theta) \rangle / [\langle \hat{X}_i(\theta)^2 \rangle \langle \hat{X}_j(\theta)^2 \rangle]^{1/2}$, 
thus minimizing the inferred quadrature variance.

Our choice of generalized quadrature
operators \cite{PhysRevA.78.060104} varies from the standard form, $\hat{X}_j(\theta)=\hat{a}_je^{-i\theta} +
\hat{a}_j^{\dagger}e^{i\theta}$ \cite{walls2010quantum}, as it does not assume a 
perfectly coherent, strong local oscillator.
Instead, it takes into account the fact that the local oscillator is 
derived, just before the measurement time, from the partially depleted and already incoherent pump mode \cite{Oberthaler-2011}. When measuring these 
quadratures the pump mode is split into two local oscillators by an atomic beam-splitter \cite{PhysRevA.78.060104} (for instance a rf $\pi/2$ pulse), in which the output is given by $\hat{b}_{\pm1} = (\hat{a}_{0} \pm\hat{a}_{\mathrm{vac}})/\sqrt{2}$, 
where $\hat{a}_{\mathrm{vac}}$ represents the vacuum entering the empty port of the beam-splitter. This is slightly different 
to the method used in Ref. \cite{Oberthaler-2011}, where an atomic three-port beam-splitter is used to measure relevant quadratures.

%

Phase accrued due to $\hat{H}_{\mathrm{el}}+\hat{H}_{\mathrm{Z}}$
leads to a drifting in the phase relation between the local oscillator and
the signal/idler modes. This means that our original quadrature choice of
$\hat{X}_{j}(\pi/4)$
and $\hat{X}_{j}(3\pi/2)$ may not measure the
optimal violation of the EPR criterion. By minimizing this criterion as a function of phase the optimal choice of quadratures becomes
$\hat{X}_{j}(\theta_0(\tau))$ and
$\hat{X}_{j}(\theta_0(\tau)+\pi/2)$, where $\theta_0(\tau)$ is the optimal local oscillator phase relative to the signal/idler modes.

\begin{figure}
\includegraphics[width=8.2cm]{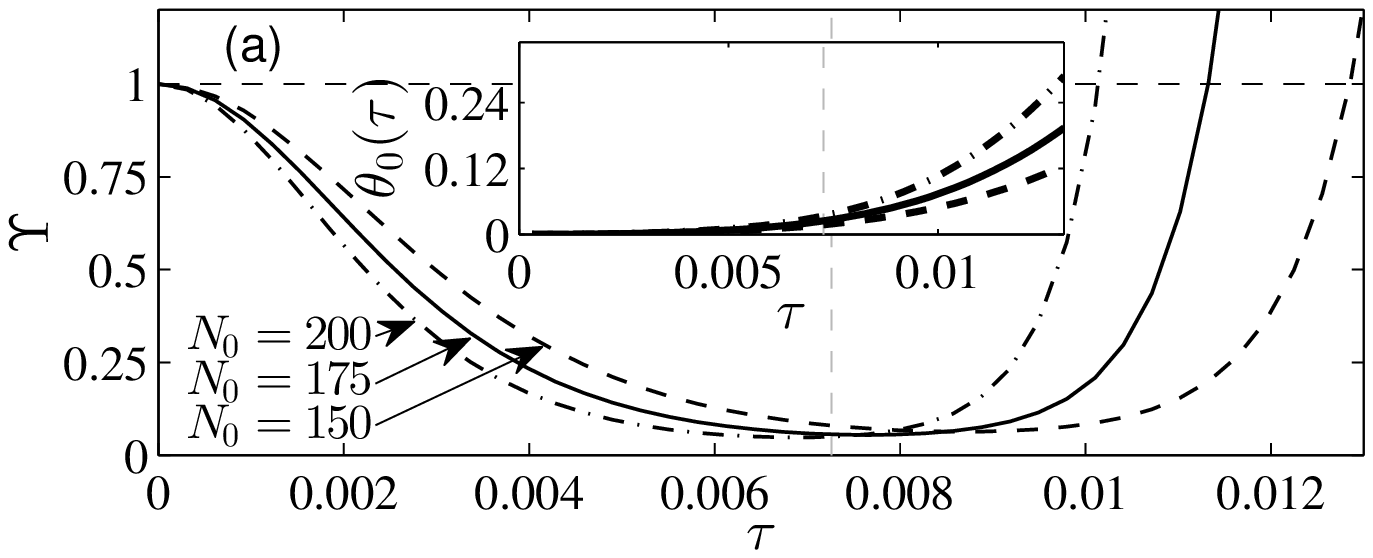}
\includegraphics[width=8.2cm]{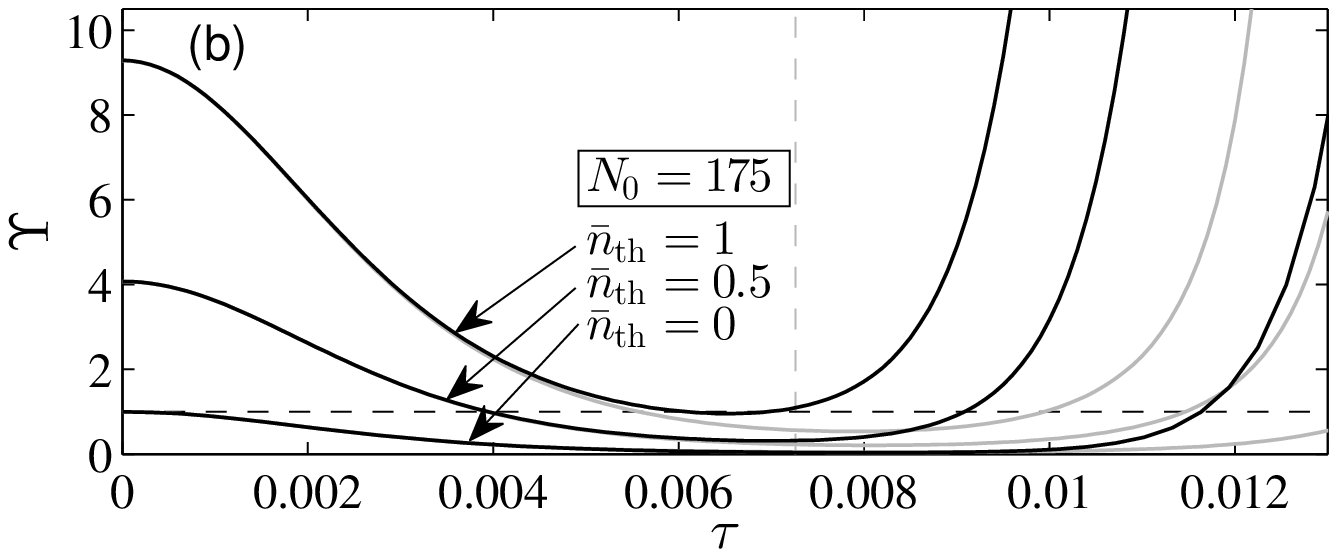}
\includegraphics[width=8.2cm]{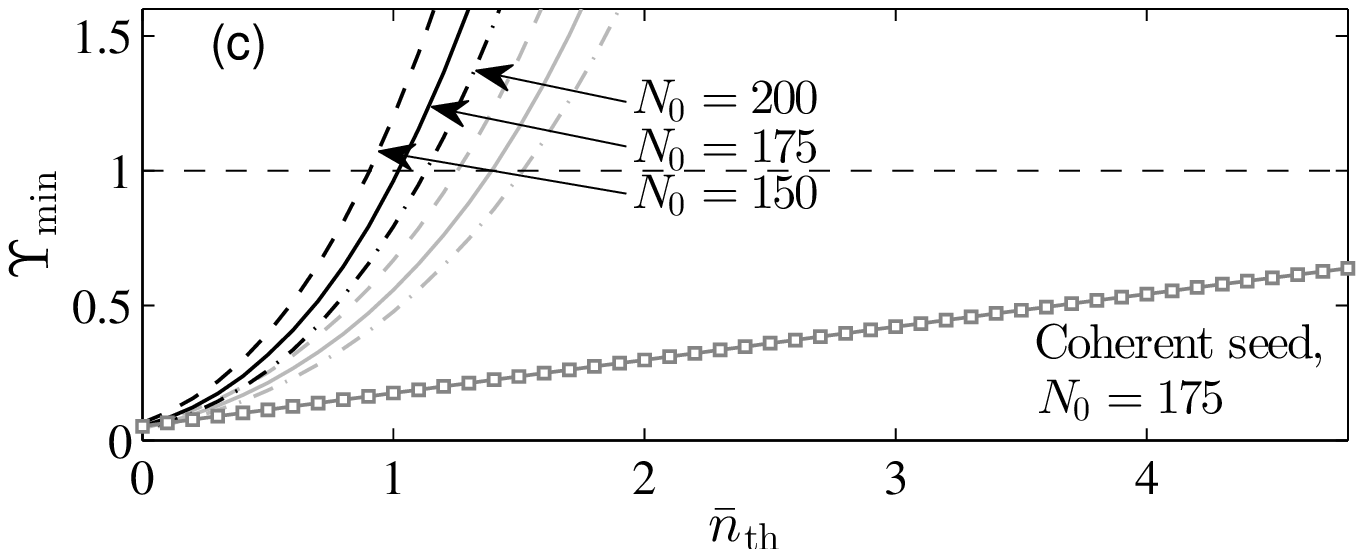}
\includegraphics[width=8.2cm]{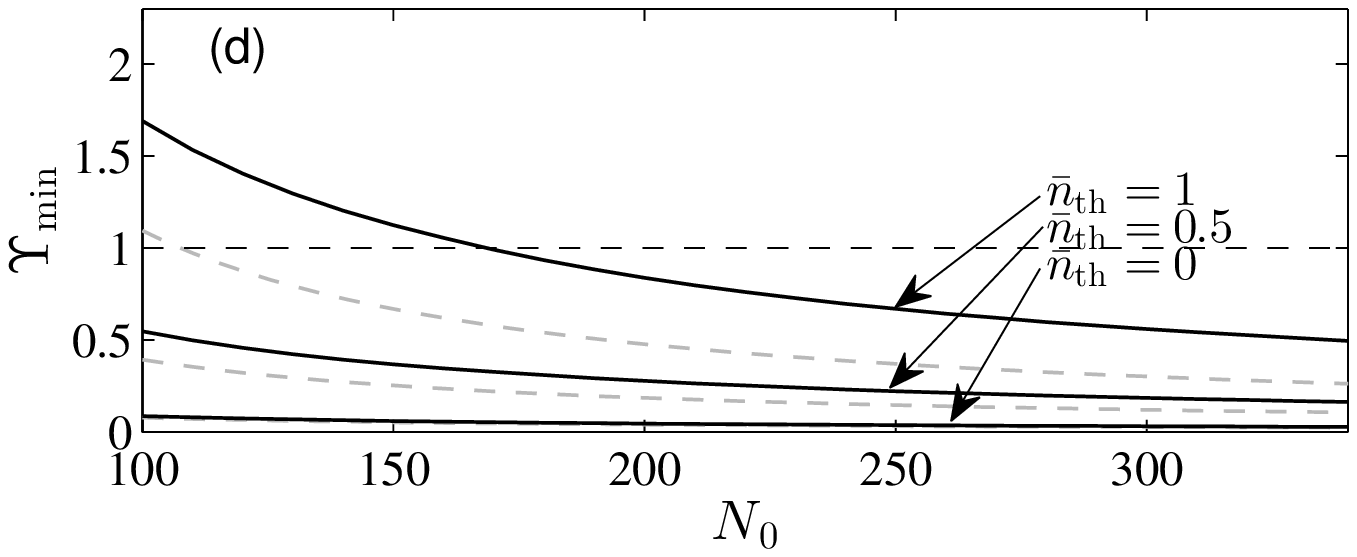}
\caption{(a) Evolution of the EPR entanglement parameter 
$\Upsilon$
for the same situation as in Fig.~\ref{fig:spin-mixing}~(a). 
The EPR criterion corresponds to 
$\Upsilon<1$ (dashed horizontal line). The inset shows 
the evolution of the optimal phase angle of the local oscillator
$\theta_0(\tau)$ for each $N_0$. 
(b) Evolution of $\Upsilon$ for thermally seeded signal/idler modes and
$N_0\!=\!175$.
The experimental measurement time $\tau'\!=\!0.0073$ is shown in (a)-(b) as a
vertical dotted line. The respective grey lines 
are the analytic predictions from the undepleted pump model.
(c) Time-optimized EPR parameter $\Upsilon_{\min}$ as a function of 
$\bar{n}_{\mathrm{th}}$, for different $N_0$. The respective grey lines 
are the analytic predictions from the undepleted pump model.
The grey line with squares shows $\Upsilon_{\min}$ for $N_0\!=\!175$, but assuming that the
seeds are in 
a coherent state 
(sharing initially the same phase as the pump mode) with average populations of
$|\alpha_{\pm1}(0)|^2=\bar{n}_{\mathrm{th}}$. 
(d) Same as in (c), but as a function of $N_0$, for three different thermal
seeds $\bar{n}_{\mathrm{th}}$.}
\label{fig:EPR}
\end{figure}

In Fig.~\ref{fig:EPR} (a) we show the results of calculation of the
phase-optimized 
EPR entanglement 
parameter $\Upsilon$ (with $\Upsilon_{-1}=\Upsilon_{1}\equiv \Upsilon$ due to 
the symmetry of the signal/idler modes) for the signal/idler modes initially in a vacuum 
state. We see that strong EPR entanglement ($\Upsilon<1$) can be achieved for
a large experimental time frame, up to $\tau\simeq0.01$; more specifically, 
we predict suppression of the optimal EPR entanglement of at least $90\%$ 
below unity for all relevant total atom numbers (ranging from $150$ to $200$) at
$\tau'=0.0073$.  Unlike the simple
undepleted pump model, which predicts $\Upsilon=\cosh^{-2}(2N_0\tau)$ 
and hence indefinite suppression of the EPR criterion \cite{walls2010quantum}, 
EPR entanglement in the full model is eventually lost due to a combination
of back-conversion
($\left|+1\right\rangle
+\left|-1\right\rangle \rightarrow\left|0\right\rangle +\left|0\right\rangle $)
and the loss of coherence in the pump mode.

Our results predict that a strong EPR violation should have
been observed if the signal and idler modes were indeed generated from an
initial vacuum state.
In light of this and the large error margin of the experimental result in Ref. \cite{Oberthaler-2011}, which thus cannot conclusively demonstrate the existence
or non-existence of EPR entanglement, we now discuss the possible presence of stray or
thermally excited 
atoms in the signal/idler modes and the effects such
seeding can have on entanglement and particularly the EPR criterion. 
The results of calculation of the EPR entanglement parameter $\Upsilon$ 
for an initial thermal seed of $\bar{n}_{\mathrm{th}}$ in both modes are
shown in 
Figs. \ref{fig:EPR} (b)-(d). We find the introduction of a
thermal seed reduces the strong correlation between the signal and idler modes,
 leading to an eventual loss of EPR entanglement. For an initial number of
atoms 
in the pump mode ranging between $150$ to $200$, EPR entanglement is lost
already for $\bar{n}_{\mathrm{th}}\simeq1$. Direct experimental detection of stray atoms
at such a low population level 
is beyond the current resolution of absorption imaging techniques
\cite{Oberthaler-2011}. More generally, our numerical results show that the maximum 
$\bar{n}_{\mathrm{th}}$ that can be tolerated while preserving the EPR entanglement scales as 
 $(\bar{n}_{\mathrm{th}})_{\mathrm{max}} \sim 0.06N_{0}^{11/20}$
in the range of $100\lesssim{N}_{0}\lesssim400$ (see Appendix A for further discussion). 
For comparison,
seeding the signal and idler modes with a coherent state
\cite{PhysRevLett.104.195303} of similar population 
does not have such a dramatic effect on EPR entanglement [see the grey 
line with squares in Fig. \ref{fig:EPR} (c)].

\begin{figure}
\includegraphics[width=8.2cm]{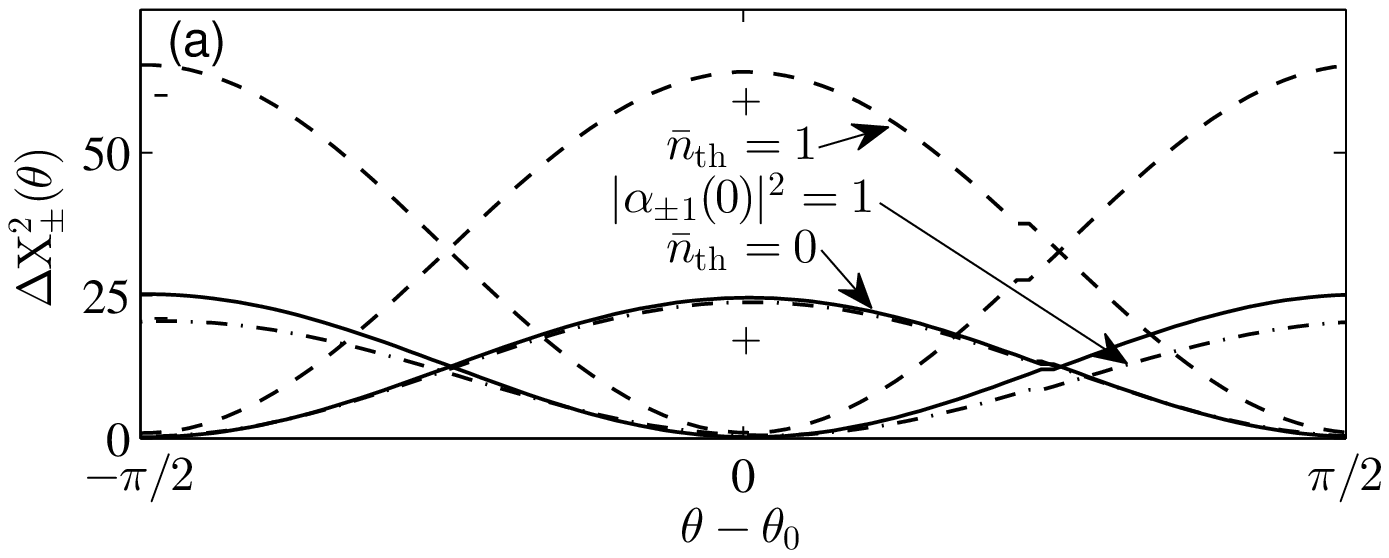}
\includegraphics[width=8.2cm]{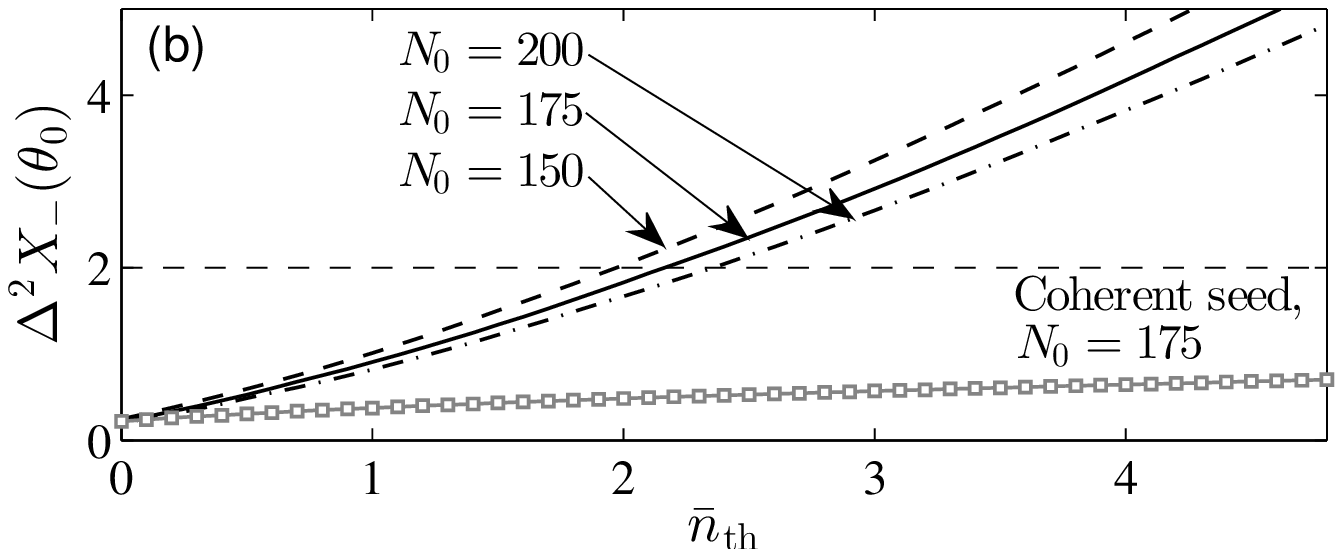}
\includegraphics[width=8.2cm]{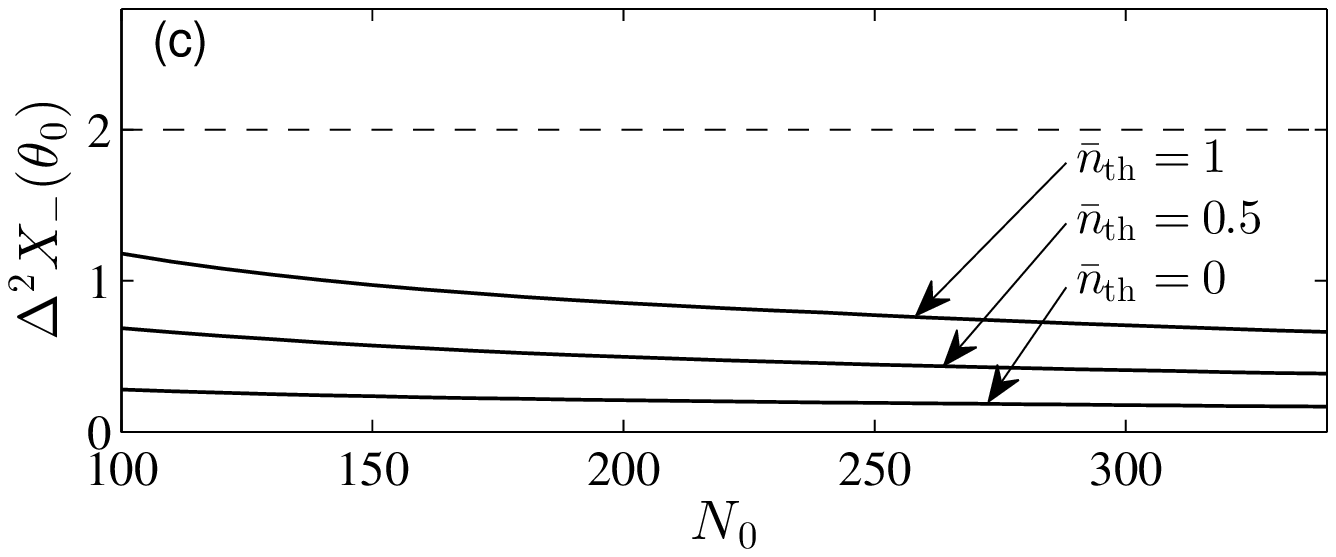}
\caption{(a) Two-mode quadrature variances
$\Delta^{2}\hat{X}_{\pm}\left(\theta\right)$ at $\tau'=0.0073$
as functions of the local oscillator phase angle $\theta-\theta_0$, 
for vacuum (solid lines) and thermally seeded (dashed lines)
signal/idler modes; $N_{0}=175$ in both cases. We also include a calculation of $\Delta^{2}\hat{X}_{-}\left(\theta\right)$ for
 comparable coherent seed (dot-dashed line), $|\alpha_{\pm1}(0)|^2= 1$, which is almost indistinguishable from the vacuum case.  
(b) Time-optimized 
minimum of $\Delta^{2}\hat{X}_{-}(\theta_0)$ as a function of $\bar{n}_{\mathrm{th}}$, for different $N_0$. The grey line with squares shows 
$\Delta^{2}\hat{X}_{-}(\theta_0)$ for $N_0 = 175$, but assuming the seeds are in a coherent state with average populations of $|\alpha_{\pm1}(0)|^2=\bar{n}_{\mathrm{th}}$.
(c) Same as in (b), but as a function of $N_0$, for different  $\bar{n}_{\mathrm{th}}$. 
\label{fig:two-mode-quadrature-variance}}
\end{figure}

\subsection{Quadrature squeezing and inseparability}

To further highlight the high sensitivity of EPR entanglement to initial
thermal noise we
contrast it with two other weaker measures of the nonclassicality of the state: 
two-mode quadrature squeezing and intermode entanglement in the sense of
inseparability, 
which were the main focus of Ref. \cite{Oberthaler-2011}.
The two-mode quadrature variances are defined as
$\hat{X}_{\pm}\left(\theta\right)=\hat{X}_{1}\left(\theta\right)\pm\hat{X}_{-1
}\left(\theta\right)$,
with $\Delta^{2}\hat{X}_{-}(\theta)<2$ corresponding to two-mode squeezing
\cite{walls2010quantum}, 
\textit{i.e.},
suppression of fluctuations below the level dictated by a minimum uncertainty
state.
We plot the results of our numerical calculations of quadrature variances in
Fig.~\ref{fig:two-mode-quadrature-variance}~(a). From these results we observe
that
the measurements
of Ref.~\cite{Oberthaler-2011} do not agree with the amplitude of the
oscillation that we find for 
an initial vacuum state (solid lines) or a small coherent seed (dot-dashed line). 
Rather they suggest the presence
of a small thermal seed of $\bar{n}_{\mathrm{th}}\simeq1$ (dashed lines), although for definitive differentiation of initial thermal or coherent populations 
further experimental measurements with reduced error margins are required. 
Further, calculation of 
the minimum of $\Delta^{2}\hat{X}_{-}$ 
[Figs. \ref{fig:two-mode-quadrature-variance} (b)-(c)] 
highlights that two-mode squeezing is preserved
for thermal seed populations up to $\bar{n}_{\mathrm{th}}\simeq 1.7$, which is consistent with our interpretation of the measurements reported in Ref. \cite{Oberthaler-2011}.

\begin{figure}
\includegraphics[width=8.3cm]{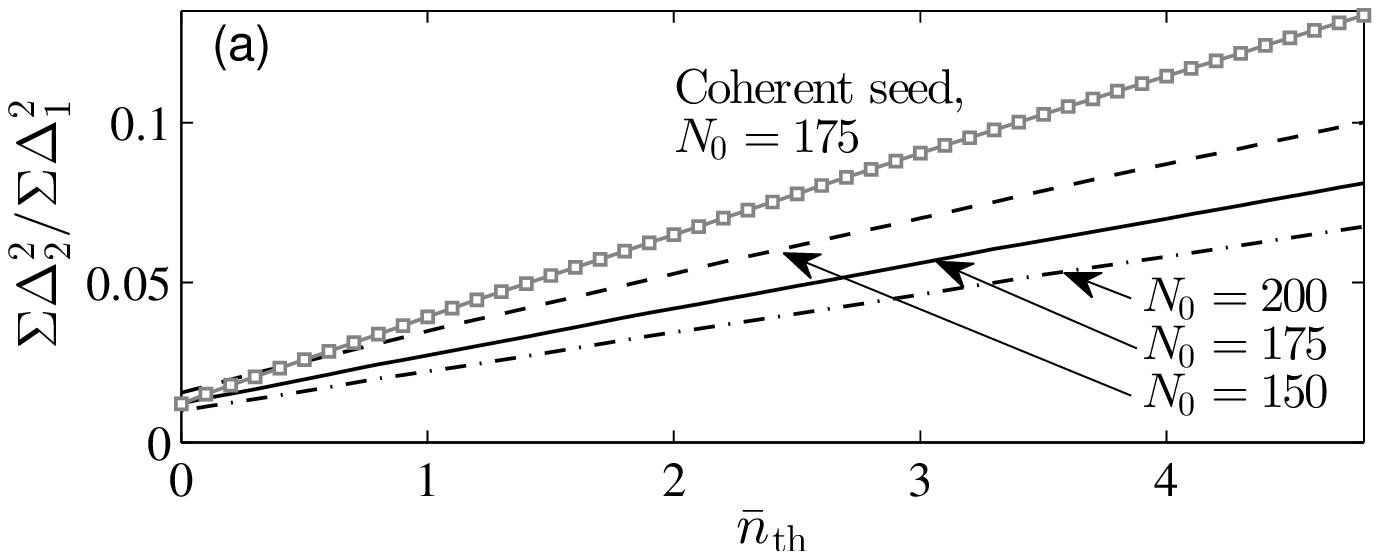}
\includegraphics[width=8.3cm]{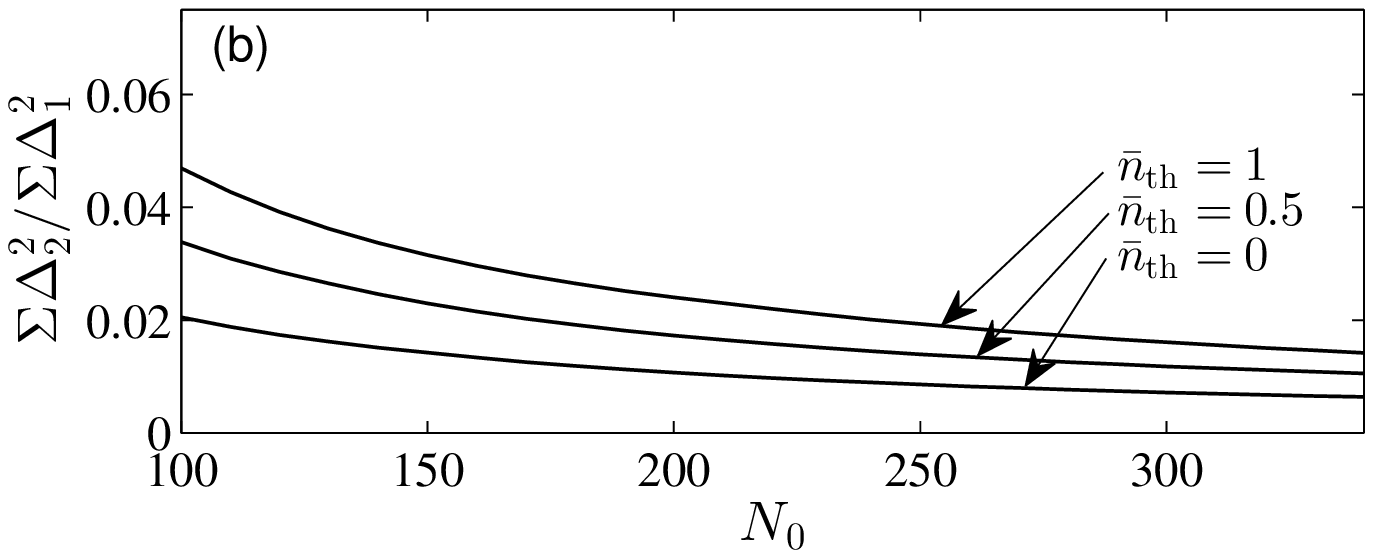}
\caption{(a) Time-optimized inseparability criterion for the
quadrature 
entangled state, quantified via 
$\sum\Delta_{2}^{2}/\sum\Delta_{1}^{2}<1$, as 
a function of  $\bar{n}_{\mathrm{th}}$, for different $N_0$. The grey line with squares shows $\sum\Delta_{2}^{2}/\sum\Delta_{1}^{2}$ for $N_0=175$, but assuming the seeds are in a coherent state 
with average populations $|\alpha_{\pm1}(0)|^2=\bar{n}_{\mathrm{th}}$.
(b) Same as in (a), but as a function of $N_0$, for different  $\bar{n}_{\mathrm{th}}$.
\label{fig:inseparability}}
\end{figure}

Next we define the sum of single-mode quadrature variances as
$\sum\Delta_{1}^{2}=2(\Delta^{2}\hat{X}_{1}+\Delta^{2}\hat{Y}_{1})$
and the sum of two-mode quadrature variances,
$\sum\Delta_{2}^{2}=\Delta^{2}\hat{X}_{-}+\Delta^{2}\hat{Y}_{+}$.
(Following the treatment of Ref. \cite{Oberthaler-2011} we calculate the
single-mode quadrature variances with the standard definition of quadratures, 
$\hat{X}_j(\theta)=\hat{a}_je^{-i\theta} + \hat{a}_j^{\dagger}e^{i\theta}$.)
Inseparability of the produced $m_{F}=\pm1$ pair-entangled state
is equivalent to
$\sum\Delta_{2}^{2}/\sum\Delta_{1}^{2}<1$ \cite{Raymer2003}.
Figures \ref{fig:inseparability} (a) and (b) demonstrate that this
measure of entanglement is far less sensitive to the presence of a
thermal seed in comparison to the stronger criterion of EPR entanglement.
Also, unlike the EPR criterion, this inseperability measure does
not significantly differentiate between coherent and thermal seeding.

\section{Summary}

In conclusion, we have demonstrated that for an initial vacuum state in the
signal/idler modes a strong suppression of the EPR criterion can be achieved 
in the parameter regime of Ref. \cite{Oberthaler-2011}, most
importantly including the experimental measurement time of $\tau'=0.0073$.
However, we also establish that the strength of EPR entanglement
depends crucially on the nature of the
initial spin-fluctuations. Specifically, we predict that
for a pump mode of initially 150 to 200 atoms, a thermal initial
seed of $\bar{n}_{\mathrm{th}}\simeq1 $ is sufficient to rule out EPR
entanglement. 
Weaker measures of entanglement, such as inseparability,
are still possible to observe as these are far more robust to thermal noise. 
This implies that spin-changing collisions may still be a good source of entanglement even in the presence of large thermal effects, 
even though we may not be able to carry through the EPR arguments that confront the completeness of quantum mechanics and 
advocate for local hidden variable theories.
Importantly, our results suggest that the measurement of this EPR criterion can serve as a sensitive probe
of the initial state which triggers the pair production process, beyond measures
employed in Ref. \cite{PhysRevLett.104.195303}.
This understanding of the sensitivity of EPR entanglement to
initial thermal noise will hopefully lead to refining of spin-mixing
experiments towards demonstration of the EPR paradox with massive particles. We expect
our findings to be also relevant to related proposals based on molecular dissociation
\cite{Dissociation-1,*Dissociation-2}, 
condensate collisions
\cite{Ferris-FWM,Jaskula:10,Cauchy-Schwarz,Zeilinger-2012}, and 
optomechanical systems \cite{Aspelmeyer-2007,*Chen-2012}.

\begin{acknowledgments}
The authors acknowledge stimulating discussions with M. Oberthaler, J.
Sabbatini, and
M. J. Davis. K.V.K. acknowledges
support by the ARC Future Fellowship award FT100100285. 
\end{acknowledgments}

%



\appendix

\section{Undepleted Pump Approximation}
\label{sec:App_A}

To invoke the undepleted pump approximation, we assume that the pump mode is initially in a coherent state with an amplitude $\alpha_0(0)=\sqrt{N_0}$ 
(which we choose to be real without loss of generality) and that it does not change with time. By additionally choosing the quadratic Zeeman effect to be phase-matched ($q=gN_{0}$), we 
can reduce the model Hamiltonian to that of optical parametric-down conversion \cite{walls2010quantum}, $\hat{H}=\hbar \chi(\hat{a}_1^{\dag}\hat{a}_{-1}^{\dag}+h.c.)$, 
in which $\chi=gN_0$. The Heisenberg equations of motion following from this are $d\hat{a}_{\pm1}/d\tau=-iN_{0}\hat{a}_{\mp1}^{\dagger}$,
where $\tau=gt$ is a dimensionless time. Solutions to these equations are given by
\begin{equation}
\hat{a}_{\pm1}(\tau) = \cosh(N_0\tau)\hat{a}_{\pm1}(0) - i\sinh(N_0\tau)\hat{a}^{\dagger}_{\mp1}(0), \label{operator_eqn_soln}
\end{equation}
which are physically valid in the short-time limit, generally corresponding to less than 10\% depletion of the pump mode occupation.

Considering specific initial states for the signal and idler modes, these solutions can be used to calculate expectation values of 
various quantum mechanical operators and observables. For example, for a thermal initial state with an equal population in both modes,
$\langle\hat{a}^{\dagger}_{1}(0)\hat{a}_{1}(0)\rangle\!=\!\langle\hat{a}^{\dagger}_{-1}(0)\hat{a}_{-1}(0)\rangle \equiv \bar{n}_{\mathrm{th}}$, 
the subsequent evolution of the mode populations is given by
\begin{equation}
\langle\hat{a}^{\dagger}_{\pm1}(\tau)\hat{a}_{\pm1}(\tau)\rangle =  
\sinh^{2}(N_0\tau)[1+2\bar{n}_{\mathrm{th}}] + \bar{n}_{\mathrm{th}},
\end{equation}
whereas the anomalous moments evolve according to
\begin{equation}
\langle\hat{a}_{\pm1}(\tau)\hat{a}_{\mp}(\tau)\rangle = -i \sinh(N_0\tau)\cosh(N_0\tau)[1+2\bar{n}_{\mathrm{th}}].
\end{equation}

\begin{widetext}
Similarly, the EPR entanglement parameter is found to be given by
\begin{equation}
\Upsilon\cong\left[\frac{\left(1+2\bar{n}_{\mathrm{th}}\right)^{2}+\frac{1}{N_{0}}\left[\left(1+2\bar{n}_{\mathrm{th}}\right)\cosh \left(2N_{0}\tau\right)-1\right]\left[2\left(1+2\bar{n}_{\mathrm{th}}\right)\cosh \left(2N_{0}\tau\right)-1\right]}{\left(1+2\bar{n}_{\mathrm{th}}\right)\cosh\left(2N_{0}\tau\right)-\frac{1}{N_{0}}\left[\left(1+2\bar{n}_{\mathrm{th}}\right)\cosh\left(2N_{0}\tau\right)-1\right]^{2}}\right]^{2},
\label{eq:Ferris_EPR}
\end{equation}
where we have assumed $N_0\gg1$.
The minimum value of this quantity (with respect to time $\tau$) gives the maximal violation of the EPR criterion,
\begin{equation}
\Upsilon_{\mathrm{min}}\cong\left[\frac{\sqrt{2N_{0}}}{\sqrt{\frac{1}{2}N_{0}}-\left(1+2\bar{n}_{\mathrm{th}}\right)-\frac{1}{2N_{0}}\left[\left(1+2\bar{n}_{\mathrm{th}}\right)^{3}-\left(\left(1+2\bar{n}_{\mathrm{th}}\right)^{2}+1\right)\sqrt{2N_{0}}\right]}-2\right]^{2},  \label{eq:Ferris_EPR_min}
\end{equation}
which is achieved at the optimal time
\begin{equation}
\tau_{\mathrm{min}}=\frac{1}{2N_{0}}\mathrm{arccosh}\left[-\frac{1}{2}\left(1+2\bar{n}_{\mathrm{th}}\right)
+\frac{1}{2}\sqrt{\left(1+2\bar{n}_{\mathrm{th}}\right)^{2}+2N_{0}}\right].
\end{equation}
\end{widetext}
From Eq.~(\ref{eq:Ferris_EPR_min}) we also determine the maximum allowable thermal population before EPR entanglement is lost. 
By numerical analysis we find a maximum seed of $(\bar{n}_{\mathrm{th}})_{\mathrm{max}}\simeq 0.05N_{0}^{2/3}$ in the range $100 \leqslant{N}_{0}\leqslant400$. We find this compares reasonably with the results of full numerical simulations, which predict $(\bar{n}_{\mathrm{th}})_{\mathrm{max}}\simeq0.06N_{0}^{11/20}$.

Furthermore we may also calculate the minimum two-mode quadrature variance,
\begin{equation}
\Delta^{2}X_{-}=2(1+2\bar{n}_{\mathrm{th}})[\cosh(2N_0\tau)-\sinh(2N_0\tau)], 
\label{min_two_mode_quad_var}
\end{equation}
and the inter-mode inseparability parameter (see main text),
\begin{equation}
\Sigma\Delta^{2}_{2}/\Sigma\Delta^{2}_{1}=1-\mathrm{tanh}(2N_{0}\tau).
\label{inter-mode_entanglement}
\end{equation}

Despite their limited applicability and the quantitative disagreement with the numerical results, the analytic predictions of the undepleted pump approximation
give useful insights into the qualitative aspects of different measures of entanglement. For example, to leading order, Eqs. (\ref{eq:Ferris_EPR}) and (\ref{min_two_mode_quad_var}) predict, respectively, quadratic and linear growth of the EPR entanglement parameter and two-mode squeezing 
with the thermal seed $\bar{n}_{\mathrm{th}}$, whereas the inter-mode inseparability, Eq.~(\ref{inter-mode_entanglement}), is insensitive to $\bar{n}_{\mathrm{th}}$. The predictions for EPR entanglement and two-mode squeezing are in qualitative agreement with the numerical results discussed in the main text, whilst we find weak linear growth with $\bar{n}_{\mathrm{th}}$ emerges for inter-mode inseparability due to depletion of the pump. These qualitative predictions highlight the lower tolerance and higher sensitivity of the EPR entanglement to thermal noise.


\begin{thebibliography}{37}
\bibitem [{coh()}]{coherent}%
  \BibitemOpen
  \href@noop {} {}\bibinfo {note} {This method can be easily implemented for
  modelling the pump mode being initially either in a pure Fock state or in a
  coherent state. For the large values of $N_0$ and relatively small time
  durations considered in this paper, the two alternatives give very similar
  results; accordingly, we restrict ourselves to presenting the results only
  for the coherent initial state.}\BibitemShut {Stop}%
\end{thebibliography}
\end{document}